\documentclass[aps,prx,showpacs,preprintnumbers,twocolumn,nofootinbib]{revtex4-2}
\usepackage{amsmath,amssymb}
\usepackage{bm}
\usepackage{tipa}
\usepackage{upgreek}
\usepackage{comment}
\usepackage{mathrsfs}
\usepackage{graphicx}
\usepackage{lipsum}
\usepackage{braket}
\usepackage{enumitem}
\usepackage{mathbbol}
\usepackage{booktabs}
\usepackage{gensymb}
\usepackage[normalem]{ulem}
\usepackage{color}
\usepackage{url}
\usepackage[colorlinks,bookmarks=true,citecolor=blue,linkcolor=red,urlcolor=blue]{hyperref}

\begin{document}

\hypersetup{breaklinks=true}

\title{A Closed Band-Projected Density Algebra Must be Girvin-MacDonald-Platzman}

 \author{Ziwei Wang and Steven H. Simon}
	\affiliation{Rudolf Peierls Centre for Theoretical Physics, Parks Road, Oxford, OX1 3PU, UK}

\begin{abstract}

The band-projected density operators in a Landau level obey the Girvin-MacDonald-Platzman (GMP) algebra, and a large amount of effort in the study of fractional Chern insulators has been directed towards approximating this algebra in a Chern band. In this paper, we prove that the GMP algebra, up to form factors, is the \textit{only} closed algebra that projected density operators can satisfy in two and three dimensions, highlighting the central place it occupies in the study of Chern bands in general.  A number of interesting corollaries follow.

\end{abstract}

\maketitle

{\bf Introduction:} In the study of the fractional quantum Hall effect, the authors of Refs.~\cite{girvin_collective-excitation_1985, girvin_magneto-roton_1986} found that projected density operators in the lowest Landau level (LLL) satisfy a specific closed algebra, which would later be known as the Girvin-MacDonald-Platzman (GMP) algebra. It was also realized that higher Landau levels (LLs) satisfy a similar algebra with different so-called form factors, and it is such algebra with generic form factors that we call the GMP algebra here.  (We present this algebra explicitly in Eqs. \ref{eq:Lie}, \ref{eq:FG} below).  With kinetic energy being suppressed in Landau levels and the interaction given by density-density terms, it is understood that the GMP algebra should capture Landau level physics completely. Later on, in the search of fractional Chern insulators (FCIs) \cite{BergholtzReview,SidFCIReview,BergholtzReview2}, i.e. systems that host fractional quantum Hall effect without the application of an external magnetic field, much effort was directed towards designing bands that resemble the Landau levels. Since LL physics is captured by the GMP algebra, it is deemed desirable to reproduce the GMP algebra in a Chern band, at least in some limits. In Ref.~\cite{parameswaran_fractional_2012}, the authors demonstrated that to reproduce the long wavelength limit of the GMP algebra, the Berry curvature should be constant in the Brillouin zone.   In Ref.~\cite{roy_band_2014}, the author found the necessary and sufficient condition for a band to satisfy the GMP algebra \textit{with LLL-like form factors} involves, besides Berry curvature, an additional condition on the quantum metric of the band, which would later be known as the ideal flatband condition. (The ideal flatband condition does not apply to GMP algebra with more general form factors.)

Admittedly, there have been no experimental systems, other than Landau levels, that realize the GMP algebra exactly, and according to Ref.~\cite{varjas_topological_2022}, the GMP algebra is impossible in a tight-binding model (i.e. finite number of degrees of freedom per unit cell). Nevertheless, how closely the Chern band reproduces the GMP algebra, based on Berry curvature and quantum geometry, is thought of as crucial to evaluating FCI candidates, with much theoretical efforts devoted in this direction \cite{parameswaran_fractional_2012, goerbig_fractional_2012, wu_zoology_2012, shankar_murthy, dobardzic_geometrical_2013, roy_band_2014, claassen_position-momentum_2015, jackson_geometric_2015, Bauer_2016, lee_band_2017,ozawa_relations_2021,mera_kahler_2021,mera_engineering_2021, ledwith_vortexability_2022}.
In Ref.~\cite{xie_fractional_2021},  experimental evidence supported the correlation between Berry-flatness and robustness of FCI (see also counter-arguments given in Refs.~\cite{simon_contrasting_2020,harper,varjas_topological_2022}). Furthermore, in Ref.~\cite{wang_exact_2021}, a mapping was developed between systems that satisfy the ideal flatband condition, which counts twisted bilayer graphene and other moir\'e materials (exactly so in the chiral limit, approximately so with more realistic parameters) as examples~\cite{ledwith_fractional_2020,ledwith_family_2022, wang_hierarchy_2022,ledwith_vortexability_2022, devakul_magic-angle_2023, dong_many-body_2023, gao_untwisting_2023}, to Landau levels, which satisfy the GMP algebra. In Ref.~\cite{wang_origin_2023}, the authors further argued that such ideal flatbands satisfy a density commutation relation that can be built up from the GMP algebra. As such, we believe systems with GMP algebra, as an idealized set-up that can be compared with and mapped to other less ideal systems, remains a highly relevant topic in the field of FCI research and beyond.

The GMP algebra was first derived for electrons in a strong magnetic field, and this is one possible closed algebra that projected density operators can obey. Though no other closed density operator algebra is known, one may nevertheless wonder if any other closed algebra exists, which may deserve equal research interest. In this paper, we show that \textit{the GMP algebra in two and three dimensions, up to form factors, are the only possible closed algebra that projected density operators may satisfy}, demonstrating the unique place the GMP algebra occupies in the study of Chern bands.

{\bf Main Result:}  For any Bloch band structure, one can define the band-projected density operator $\bar \rho({\bm q}) = {\cal P} e^{i {\bm q} \cdot{\bm r}}  {\cal P}$, where ${\cal P}$ projects to a single band. Suppose the algebra of these operators close
\begin{equation}
 [\bar \rho({\bm q}_1), 
 \bar \rho({\bm q}_2)] = f({\bm q}_1,{\bm q}_2) \,\, \bar \rho({\bm{q}_1+ \bm{q}_2})   \label{eq:Lie}  
\end{equation}
thus defining a Lie algebra.    We show that in two and three dimensions, $f$ can only have the form
\begin{equation}\label{eq:FG}
f({\bm{q}_1,\bm{q}_2}) = 2i\frac{F({\bm q}_1) F({\bm q}_2)}{ F({\bm q}_1 + {\bm q}_2)} \sin\left(\frac{1}{2}\bm{\Omega} \cdot ({\bm q}_1 \times {\bm q}_2)\right)
\end{equation}
for some function $F$ that satisfies $F(-\bm{q}) = F^*(\bm{q})$ and a fixed real vector $\bm \Omega$. Eqs.~\ref{eq:Lie}, \ref{eq:FG} are known as the GMP algebra, $F$ is known as the form factor, and $\bm \Omega$ will turn out to be the Berry curvature (a vector in three dimensions, and a vector normal to the plane in two dimensions). 

{\bf Main Proof:}
From the definition $\bar{\rho}_{\bm{q}} = {\cal P}e^{i \bm{q} \cdot \bm{r}}{\cal P}$, we can write the projected density operator in Bloch basis as
\begin{equation}
    \bar{\rho}_{\bm{q}} = \int_{\text{BZ}} d^2\bm{k} \braket{u_{\bm{k} + \bm{q}}|u_{\bm{k}}}\ket{\bm{k} + \bm{q}}\bra{\bm{k}}
    \label{eq:rhodef}
\end{equation}
where $\ket{\bm{k}}$ denotes the Bloch state with momentum $\bm{k}$ and $\ket{u_{\bm{k}}} = e^{-i \bm{k} \cdot \bm{r}} \ket{\bm{k}}$ is the periodic part of the Bloch state, and the integration is over the Brillouin zone (BZ). Since we will only consider a single Bloch band, the band index has been suppressed. We will choose the gauge such that the Bloch states $\ket{\bm{k}}$ vary smoothly with $\bm{k}$, but in general they do not have Brillouin zone periodicity. 

Using the orthogonality of Bloch states, we have
\begin{align} \label{eq:rhorho}
    & \bar{\rho}_{\bm{q}_1} \bar{\rho}_{\bm{q}_2} =  \\  & \int_{\text{BZ}} d^2\bm{k} \braket{u_{\bm{k} + \bm{q_1} + \bm{q}_2}|u_{\bm{k} + \bm{q}_2}}\braket{u_{\bm{k} + \bm{q}_2}|u_{\bm{k}}} \ket{\bm{k} + \bm{q}_1 + \bm{q}_2}\bra{\bm{k}}~~. \nonumber
\end{align}

We then define
\begin{equation}\label{def_ft}
    \tilde{f}(\bm{q}_1, \bm{q}_2; \bm{k})= \frac{\braket{u_{\bm{k} + \bm{q_1} + \bm{q}_2}|u_{\bm{k} + \bm{q}_2}}\braket{u_{\bm{k} + \bm{q}_2}|u_{\bm{k}}}}{ \braket{u_{\bm{k} + \bm{q}_1 + \bm{q}_2}|u_{\bm{k}}}}
\end{equation}
and
\begin{equation}\label{def_f}
f({\bm{q}_1,\bm{q}_2})_{\bm k} = \tilde f(\bm{q}_1,\bm{q}
_2; \bm{k}) - \tilde f({\bm{q}_2,\bm{q}_1; \bm{k}}) ~~. 
\end{equation}

Plugging Eqs.~\ref{eq:rhorho} and \ref{eq:rhodef} into Eq.~\ref{eq:Lie} it is easy to establish that in order for the $\bar{\rho}_{\bm{q}}$ operators to be closed under commutation it must be the case that $f({\bm{q}_1,\bm{q}_2})_{\bm k} $ is $\bm k$-independent, so that
$$
f({\bm{q}_1,\bm{q}_2})_{\bm k} = 
f({\bm q_1,\bm q_2})~~~.
$$

\textit{The aim of this work is to show that if the function $f({\bm{q}_1,\bm{q}_2})_{\bm k}$, as defined in Eq.\ref{def_ft} and \ref{def_f}, is independent of $\bm{k}$, it must take the form of Eq.~\ref{eq:FG}.} 

We first consider the expression
$
f({\bm q} - {\bm \epsilon}, {\bm{\epsilon}})_{\bm k}
$
for small $\bm{\epsilon}$.  Note that it is zero for $\bm{\epsilon} = 0$.  Using the definitions Eqs.~\ref{def_ft} and \ref{def_f}, we expand to first order in $\bm{\epsilon}$, and we find
$$
 f({\bm q} - {\bm \epsilon}, {\bm{\epsilon}})_{\bm k}  = -{\bm \epsilon} \cdot {\bm G}(\bm q)_{\bm k}
$$
with 
\begin{equation}\label{eq:def_G}
    {\bm G}(\bm q)_{\bm k} = i  {\bm {\mathcal A}}(\bm{k}+\bm{q}) - i {\bm {\mathcal A}}(\bm{k}) + \bm{\nabla}_{\bm k}  \log \langle u_{\bm{k} + \bm{q}} | u_{\bm k} \rangle 
\end{equation}
where we have defined the Berry connection as
$$
\bm{\mathcal A}(\bm k) =  -i \langle u_{\bm k} |\bm{\nabla}_{\bm k} | u_{\bm k} \rangle~~~~~.
$$
For the moment, we have assumed that $\langle u_{\bm{k} + \bm{q}} | u_{\bm k} \rangle \neq 0$ so that its logarithm is well-defined. We will revisit the issue of possibly singularities later on. 

In Ref.~\cite{parameswaran_fractional_2012}, the authors found that, in 2-dimensions,
\begin{equation*}
   [\bar \rho_{{\bm q}_1}, \bar \rho_{{\bm q}_2}] |{\bm k}\rangle \approx i \Omega(\bm{k}) \hat{z}\cdot ({\bm{q}_1 \times \bm{q}_2)} \bar \rho_{\bm{q}_1 + \bm{q}_2} | {\bm{k}} \rangle
\end{equation*}
where $\Omega(\bm{k})$ is the Berry curvature at $\bm{k}$, in the limits where $\bm{q}_1$ and $\bm{q}_2$ are small.  This immediately implies that for the projected density operator to be closed, Berry curvature must be constant in the Brillouin zone. The result also obviously generalizes to 3-dimensions, 
\begin{equation}\label{eq:small_q}
   [\bar \rho_{{\bm q}_1}, \bar \rho_{{\bm q}_2}] |{\bm k}\rangle \approx i \bm{\Omega}(\bm{k})\cdot ({\bm{q}_1 \times \bm{q}_2)} \bar \rho_{\bm{q}_1 + \bm{q}_2} | {\bm{k}} \rangle
\end{equation}
where $\bm{\Omega}(\bm{k}) = \bm{\nabla}_{\bm{k}} \times {\bm {\mathcal A}}(\bm{k})$ is the Berry curvature vector.   For constant Berry curvature, we can write
$$
{\bm {\mathcal A}}(\bm k)  = \frac{1}{2}{\bm \Omega} \times {\bm k} + \bm{\nabla}_{\bm k} \phi(\bm k)  
$$
for some unknown real function $\phi$.  

Given this form, we have
 \begin{align} \label{eq:gc}
  {\bm G}(\bm q)_{\bm k}  - & \frac{i}{2}{\bm \Omega} \times {\bm q}  = \\ 
 &\bm{\nabla}_{\bm k} \left[i \phi({\bm{k}+\bm{q}}) - i \phi({\bm k}) + \log   \langle u_{\bm{k} + \bm{q}} | u_{\bm k} \rangle \right] \nonumber
 \end{align}

In order for our algebra to close, $f({\bm{q}_1,\bm{q}_2})_{\bm k} $  must be independent of $\bm k$, which means the left-hand-side of Eq.~\ref{eq:gc} is independent of $\bm k$. This implies
\begin{equation} \label{eq:ck}
i \phi({\bm{k}+\bm{q}}) - i \phi({\bm k}) + \log   \langle u_{\bm{k} + \bm{q}} | u_{\bm k} \rangle  =  {\bm k} \cdot {\bm W}(\bm q)    + Y(\bm q)
\end{equation}
for some unknown vector function $\bm W$ and scalar function $Y$.  Thus we have 
\begin{equation}
\langle u_{\bm{k} + \bm{q}} | u_{\bm k} \rangle  = e^{-i \phi({\bm{k}+\bm{q}}) + i \phi({\bm k}) +   {\bm k} \cdot {\bm W}(\bm q)   + Y(\bm q) } ~~~~. \label{eq:inner}
\end{equation}
We then substitute this result back into the definition of $\tilde f$  (Eq.~\ref{def_ft}) and note that all the factors of $\phi$ immediately cancel.  We then have 
\begin{eqnarray} 
& & \tilde{f}(\bm{q}_1, \bm{q}_2; \bm{k}) = \label{eq:tfr} \\ & &    e^{ (\bm{k}+\bm{q}_1) \cdot {\bm W}({\bm q}_2)  + {\bm k} \cdot {\bm W}({\bm q}_1)- {\bm k} \cdot \bm W(\bm{q}_1 + \bm{q}_2) + Y({\bm q}_1) + Y({\bm q}_2) - Y({\bm q}_1 + {\bm q}_2) }  \nonumber
\end{eqnarray}
Here we recognize $e^Y$ as being (part of) the form factor $F$.   Now we demand $f({\bm{q}_1,\bm{q}_2})_{\bm k} $ in Eq.~\ref{def_f}
be $\bm k$ independent.  This then requires that
\begin{equation}\label{eq:cauchy}
{\bm W}({\bm q}_2)  + {\bm W}({\bm q}_1) - \bm W(\bm{q}_1 + \bm{q}_2)    = 0
\end{equation}
which is Cauchy's functional equation.   Given continuity of the function $\bm W$, the only solution is~\cite{functional}
\begin{equation}\label{eq:cauchy_solution}
    {\bm W}({\bm q}) = {\underline M} {\bm q}
\end{equation}
where $\underline M$ is some fixed matrix acting on $\bm q$.    We then obtain
$$
f({\bm{q}_1,\bm{q}_2}) = \frac{\tilde F({\bm q}_1) \tilde F({\bm q}_2)}{ \tilde F({\bm q}_1 + {\bm q}_2)} \left[ e^{{\bm q_1} \cdot {\underline M} {\bm q_2}} - e^{{\bm q_2} \cdot {\underline M} {\bm q_1}}  \right]  
$$
with $\tilde F = e^Y$.  The symmetric part of $\underline M$ can be absorbed into the form factors $F$ by defining 
\begin{equation} 
F(\bm{q}) = \tilde F(\bm{q})e^{-\frac{1}{2}\bm{q} \cdot {\underline M}\bm{q}} \label{eq:Fdef}
\end{equation}
we then have 
$$
f({\bm{q}_1,\bm{q}_2}) = 2\frac{F({\bm q}_1) F({\bm q}_2)}{ F({\bm q}_1 + {\bm q}_2)} \sinh[\frac{1}{2} ({\bm q}_1 \cdot {\underline M} {\bm q}_2 - {\bm q}_2 \cdot {\underline M} {\bm q}_1 ) ]
$$
which could also be written  in the form of Eq.~\ref{eq:FG} for some unknown constant vector $\bf \Omega$.

Given the form of Eq.~\ref{eq:inner}, as $\bm q \rightarrow \bm 0$, we have $Y(\bm 0) \rightarrow 0$.  Thus for small $\bm q_1, \bm q_2$, the prefactor $\frac{F({\bm q}_1) F({\bm q}_2)}{ F({\bm q}_1 + {\bm q}_2)}$ becomes unity.   Now, matching the small $\bm{q}$ limit in Eq.~\ref{eq:small_q}, we identify the unknown constant vector $\bm{\Omega}$ as the Berry curvature vector (in 2-dimension, it simply reduces to $\bm \Omega = \Omega \hat{z}$, where $\Omega$ is the scalar Berry curvature). That $F({\bm q}) = F^*({ -\bm{q}})$ can be easily established from its construction, as elaborated in Appendix~\ref{form_factor}, and we have reached the desired result.

{\bf Singularities:}  Now we revisit the issue of possible singular behaviour in Eq.~\ref{eq:def_G} due to vanishing $\langle u_{\bm{k} + \bm{q}} | u_{\bm k} \rangle$, which also leads to divergences in $f$. Since overlaps are smooth, we assume  $\langle u_{\bm{k} + \bm{q}} | u_{\bm k} \rangle$ only vanishes on lower dimensional sub-manifolds. One might worry that in the space of $\bm{k,q}$, the sub-manifold where $\langle u_{\bm{k} + \bm{q}} | u_{\bm k} \rangle$ vanishes could separate the space into disconnected regions.  If this is true, Eq.~\ref{eq:ck} could have different solutions in the different regions. 

Consider a region free of singularity. From Eq.~\ref{eq:inner}, we obtain
\begin{equation} \langle u_{\bm{k}} | u_{\bm k  + \bm{q}} \rangle  = e^{-i \phi({\bm{k}}) + i \phi({\bm k + \bm q}) +  ({\bm k + \bm q})  \cdot {\bm W}(- \bm q)   + Y(- \bm q)}~~. \label{eq:inner2}
\end{equation}
But $\langle u_{\bm{k}} | u_{\bm k + \bm q} \rangle = \langle u_{\bm{k} + \bm{q}} | u_{\bm k} \rangle^* $, so comparing Eqs.~\ref{eq:inner} and \ref{eq:inner2} we have
\begin{equation}\label{eq:WY}
e^{({\bm k + \bm q}) \cdot {\bm W}(- \bm q)     + Y(- \bm q)} = e^{{\bm k}  \cdot {\bm W}^*(\bm q)   + Y^*(\bm q)}
\end{equation}
Comparing the $\bm{k}$ dependence, we have
$\bm{W}(-\bm{q}) = \bm{W}^*(\bm{q})$.   Knowing that 
${\bm W}({\bm q}) = {\underline M} {\bm q}
$
we find ${\underline M}^* = -{\underline M}$, i.e. $\bm{W}$ is purely imaginary. Thus
$$
|\langle u_{\bm{k} + \bm{q}} | u_{\bm k} \rangle| = |e^{Y(\bm{q})}|
$$
is independent of $\bm{k}$.

This means, for a fixed $\bm{q}$, in each region where $\langle u_{\bm{k} + \bm{q}} | u_{\bm k} \rangle$ does not vanish, $|\langle u_{\bm{k} + \bm{q}} | u_{\bm k} \rangle|$ must be constant. Given $\langle u_{\bm{k} + \bm{q}} | u_{\bm k} \rangle$ is continuous, we conclude that for a fixed $\bm{q}$, if $\langle u_{\bm{k} + \bm{q}} | u_{\bm k} \rangle$ is non-zero at any $\bm{k}$, it is non-zero for all $\bm{k}$. That is, the only possible singular behaviour in Eq.~\ref{eq:ck} is where the logarithm diverges for a sub-manifold of $\bm{q}$ and all $\bm{k}$

In the main proof above we invoked the solution to Cauchy's functional equation. The same solution (Eq.~\ref{eq:cauchy_solution}) must hold  even if one allows for discontinuities or singularities in $\bm W(\bm q)$ on a set of measure zero~\cite{Cauchy2}.    Thus, the main proof presented above holds even with singularities. 

{\bf Further Results: } (1) We can also consider closed multiplication algebras
$$
\bar{\rho}_{\bm{q}_1}\bar{\rho}_{\bm{q}_2} = \tilde f(\bm{q}_1,\bm{q}_2) \bar{\rho}_{\bm{q}_1 + \bm{q}_2} ~~~. 
$$
Obviously this implies a closed commutator algebra.

Landau levels do indeed satisfy such a closed multiplication algebra as well as the closed commutation algebra.  This is not a coincidence.  From the form of Eq.~\ref{eq:tfr} we also obtain the most general form of $\tilde f$, independent of $\bm k$. That is, projected density operators that are closed under commutation are also closed under multiplication.    

(2) The quantum geometry tensor is defined as
\begin{equation} \label{eq:Qab}
    Q^{ab}(\bm{k}) = \braket{\partial_{k_a} u_{\bm{k}} | Q_{\bm{k}} |\partial_{k_b} u_{\bm{k}}}
\end{equation}
where $Q_{\bm{k}} = 1 - \ket{u_{\bm{k}}}\bra{u_{\bm{k}}}$. Its symmetric part is the quantum metric tensor while its anti-symmetric part is the Berry curvature.    Assuming a GMP algebra, it is easy to use Eq.~\ref{eq:inner}
to establish that the $Q_{ab}({\bm k})$ must also be independent of $\bm k$.  (This result was known for GMP algebra with LLL-like form factors previously\cite{roy_band_2014} but, to our knowledge, not more generally.)  We emphasize that the converse is not true:  a band with both constant Berry curvature and constant quantum metric tensor is not necessarily GMP.  An example of this is given explicitly in Appendix~\ref{example}. 

{\bf Concluding Remarks:} We recall that Girvin-MacDonald-Platzman algebra is a closed commutation algebra obeyed by projected density operators that naturally arise from the study of Landau levels. In this paper, we prove that \textit{the GMP algebra is the only possible form of closed algebra that projected density operator of a band can take in 2-dimensions and 3-dimensions}, up to the choice of form factors. We believe this result is of mathematical interest on its own. More relevant to condensed matter research, this result demonstrates the unique place the GMP algebra occupies in the study of Chern bands, and provides some further theoretical motivation for designing Chern bands similar to Landau levels. We further remark that while in the literature the greatest focus has usually been on designing LLL-like Chern bands, for example, the formulation of ideal flatband condition, an alternative perspective would be to design Chern bands with (approximate) closed algebra, which would favor the use of somewhat different criteria, as GMP algebra with generic form factors do not obey the ideal flat band condition (see also Refs.~\cite{fujimoto_higher_2024, liu_theory_2024}). We emphasize that the quantum metric tensor only reflects the long wavelength properties of the system, as further elaborated in Appendix~\ref{example}, where a band with flat quantum geometry tensor but without closed density algebra is explicitly constructed. We note, in addition, that quantum geometry alone is not enough to determine what type of ground state a system is likely to exhibit --- the details of the interaction are clearly just as important~\cite{simon_contrasting_2020,ledwith_vortexability_2022,harper}.

{\bf Acknowledgements: }Z.W. acknowledges funding from Leverhulme Trust International Professorship grant (number LIP-202-014). S.H.S. acknowledges support from EPSRC grant EP/X030881/1.

\begin{appendix}
\section{Properties of the form factor}\label{form_factor}
Starting with Eq.~\ref{eq:WY} and comparing the $\bm{k}$-independent component on both sides, we find that
$
    Y(-\bm{q})  = Y^*(\bm{q}) -  \bm{q} \cdot \bm{W}(-\bm{q})   = Y^*(\bm{q}) + \bm{q} \cdot {\underline M}\bm{q} 
$.  From Eq.~\ref{eq:Fdef} we have
$$F(\bm{q}) = e^{Y(\bm{q})  -\frac{1}{2}\bm{q} \cdot {\underline M}\bm{q}}$$
This means 
\begin{align*}
    F(-\bm{q}) & = e^{{Y(-\bm{q})  -\frac{1}{2}\bm{q} \cdot {\underline M}\bm{q}}}  = e^{ Y^*(\bm{q}) +  \frac{1}{2}\bm{q} \cdot {\underline M}\bm{q}}\\
    & = e^{ Y^*(\bm{q}) -  \frac{1}{2}\bm{q} \cdot {\underline M}^*\bm{q}} = F^*(\bm{q})
\end{align*}
where we have used $\underline{M}^* = - \underline{M}$.
As such, we have established $F(-\bm{q}) = F^*(\bm{q})$.

\section{Flat quantum geometry without closed algebra}\label{example}

Consider Landau levels on a torus or infinite plane. We choose unit cell dimensions $a_x$, $a_y$ such that $a_xa_y$ encloses one magnetic flux quantum. Define $\bm{a}_1 = (a_x, 0)$ and $\bm{a}_2 = (0, a_y)$, and the magnetic translation by $\bm{R}$ is denoted by $t_{\bm{R}}$ with $t_{-\bm{R}} = t^\dagger_{{\bm{R}}}$. The usual Landau level Hamiltonian is denoted $H_0 = \hbar \omega_c (a^\dagger a + \frac{1}{2})$, where $a^\dagger$ is the Landau level raising operator and $\omega_c$ is the cyclotron frequency. We use units where the magnetic length $l_B \equiv 1$ and $\hbar \equiv 1$.

We consider a system with a three-state pseudo-spin degree of freedom.   Denoting $t \equiv t_{2\bm{a}_1}$ for simplicity, we consider the following Hamiltonian 
\begin{widetext}
\begin{equation*}
    \begin{pmatrix}
        H_0 + \frac{1}{2}(t + t^\dagger)\epsilon_0 & \frac{i}{\sqrt{2}}(t - t^\dagger)a\epsilon_0 & \left( -1 + \frac{1}{2}(t + t^\dagger)\right)\frac{1}{\sqrt{2}}a^2\epsilon_0  \\
    \frac{i}{\sqrt{2}}(t - t^\dagger)a^\dagger\epsilon_0 & H_0 -  \omega_c - (1 + t + t^\dagger)\epsilon_0 & \frac{i}{2}(t - t^\dagger)a\epsilon_0 \\
    \left( -1 + \frac{1}{2}(t + t^\dagger)\right)\frac{1}{\sqrt{2}}
    (a^\dagger)^2\epsilon_0 & \frac{i}{2}(t - t^\dagger)a^\dagger\epsilon_0 & H_0 - 2\omega_c + \frac{1}{2}(t + t^\dagger)\epsilon_0
    \end{pmatrix}
\end{equation*}
\end{widetext}
This Hamiltonian has a flat band at $E = \frac{1}{2}\omega_c - 3\epsilon_0$, with the periodic parts of its wavefunctions given by (assuming $\omega_c$ and $\epsilon_0$ are chosen to avoid accidental degeneracy)
\begin{equation*}
     u_{\bm{k}}(\bm{r}) = \begin{pmatrix} \sin(k_xa_x)u_{0{\bm{k}}}(\bm{r})/\sqrt{2}\\ \cos(k_xa_x)u_{1{\bm{k}}}(\bm{r}) \\ \sin(k_xa_x)u_{2{\bm{k}}}(\bm{r})/ \sqrt{2}\end{pmatrix}
\end{equation*}
where $u_{n,\bm{k}}$ is the periodic part of the usual $n$-th Landau level wavefunction.  (In the context of Landau levels, we understand ``periodic" as invariant under \textit{magnetic} translation operators.) We can show that (using the same convention as Ref.~\cite{ozawa_relations_2021})
\begin{align}\label{eq:3bandoverlap}\notag
    \braket{u_{\bm{k} + \bm{q}}|u_{\bm{k}}} = & e^{- \bm{q}^2/4} e^{iq_x(k_y + q_y/2)}\left[ \cos(q_x a_x)\left( 1 - \frac{\bm{q}^2}{2}\right) \right. \\ & + \left. \frac{1}{16}\sin\left((k_x + q_x)a_x\right)\sin(k_xa_x) \bm{q}^4\right]
\end{align}
where we have used the fact that $  \braket{u_{n,\bm{k} + \bm{q}}|u_{n,\bm{k}}} =  e^{- \bm{q}^2/4} e^{iq_x(k_y + q_y/2)} L_n(q^2/2)$ with $L_n$ being the Laguerre polynomial. 

 It can be verified straightforwardly that the quantum geometry tensor (Eq.~\ref{eq:Qab}) is $\bm{k}$-independent for the band constructed above, given by 
\begin{equation*}
    \underline{Q} = \begin{pmatrix}
        5/2 & i/2 \\
        -i/2 & 3/2
    \end{pmatrix}
\end{equation*}
This means that both the Berry curvature and the quantum metric tensor are flat. However, due to the non-trivial $\bm{k}$-dependence in the $\bm{q}^4$ term in Eq.~\ref{eq:3bandoverlap}, $f(\bm{q}_1, \bm{q}_2)_{\bm{k}}$ is not $\bm{k}$-independent, and the density algebra is not closed. This example shows that flat quantum geometry tensor is not a sufficient condition for closed density algebra, and it reflects that quantum geometry is a long wavelength behaviour which does not fix the behaviour of the system at all momentum scales. (Unless the quantum metric tensor satisfies a more restrictive condition on its determinant, which leads to GMP algebra with LLL-like form factors at all momentum scales~\cite{roy_band_2014}. This special case does not apply to GMP algebra with general form factors.)

Suppose we consider the following Hamiltonian instead
\begin{equation*}
    \begin{pmatrix}
    H_0  +    (t + t^\dagger)\epsilon_0  & i(t - t^\dagger)\epsilon_0 \\
        i(t - t^\dagger)\epsilon_0 &H_0 -(t + t^\dagger)\epsilon_0 
    \end{pmatrix}
\end{equation*}
It has a flat band at $E = \frac{3}{2} \omega_c - 2\epsilon_0$, with the periodic parts of its wavefunctions given by (assuming $4\epsilon \neq n \omega_c$ for any integer $n$ to avoid accidental degeneracy)
\begin{equation*}
    \tilde{u}_{\bm{k}}(\bm{r}) = \begin{pmatrix} \sin(k_xa_x)u_{1{\bm{k}}}(\bm{r})\\ \cos(k_xa_x)u_{1{\bm{k}}}(\bm{r})\end{pmatrix} 
\end{equation*}
The overlaps are given by
\begin{equation*}
     \braket{\tilde{u}_{\bm{k} + \bm{q}}|\tilde{u}_{\bm{k}}} = e^{- \bm{q}^2/4} e^{iq_x(k_y + q_y/2)} \cos(q_x a_x)\left( 1 - \frac{\bm{q}^2}{2}\right)
\end{equation*}
which is the same as Eq.~\ref{eq:3bandoverlap} but without the $\bm{q}^4$ term. The quantum geometry tensor is identical with the previous example, but the generalized GMP algebra will be obeyed. This shows that knowledge about quantum geometry tensor is generally insufficient for diagnosing general whether the GMP algebra holds.

\end{appendix}

\bibliography{references.bib}

\end{document}